\documentclass[aps,reprint,prl,10pt,twocolumn,showkeys,nofootinbib,showpacs,superscriptaddress]{revtex4-1}

\usepackage{graphicx}
\usepackage{url}
\usepackage{amsmath}
\usepackage{amssymb}
\usepackage{bm}
\usepackage{dsfont}
\usepackage{graphicx}
\usepackage{subfigure}
\usepackage{dcolumn}
\usepackage{braket}
\usepackage{bm}
\usepackage{bbm}
\usepackage{color}
\usepackage{pdfpages}
\usepackage{pgffor}

\newcommand{\vect}[1]{\mathbf{#1}}
\newcommand{\Li}{$^{6}\mathrm{Li}\;$}

\newcommand{\affcua}{MIT-Harvard Center for Ultracold Atoms, Research Laboratory of Electronics, and Department of Physics,
Massachusetts Institute of Technology, Cambridge, Massachusetts 02139, USA}
\newcommand{\affcamb}{Cavendish Laboratory, University of Cambridge, J. J. Thomson Avenue, Cambridge CB3 0HE, United Kingdom}
\newcommand{\affens}{Laboratoire Kastler Brossel, CNRS, ENS-PSL Research University, UPMC-Sorbonne Universit\'{e}s and Coll\`{e}ge de France, Paris, France}

\begin{document}

\title{Homogeneous Atomic Fermi Gases}

\author{Biswaroop Mukherjee}
\affiliation{\affcua} 
\author{Zhenjie Yan}
\affiliation{\affcua} 
\author{Parth B. Patel}
\affiliation{\affcua} 
\author{\\Zoran Hadzibabic}
\affiliation{\affcua} 
\affiliation{\affcamb} 
 \author{Tarik Yefsah}
\affiliation{\affcua} 
 \affiliation{\affens} 
 \author{Julian Struck}
 \affiliation{\affcua} 
 \author{Martin W. Zwierlein}
 \affiliation{\affcua}

\begin{abstract}

We report on the creation of homogeneous Fermi gases of ultracold atoms in a uniform potential. In the momentum distribution of a spin-polarized gas, we observe the emergence of the Fermi surface and the saturated occupation of one particle per momentum state. This directly confirms Pauli blocking in momentum space. For the spin-balanced unitary Fermi gas, we observe spatially uniform pair condensates. For thermodynamic measurements, we introduce a hybrid potential that is harmonic in one dimension and uniform in the other two. The spatially resolved compressibility reveals the superfluid transition in a spin-balanced Fermi gas, saturation in a fully polarized Fermi gas, and strong attraction in the polaronic regime of a partially polarized Fermi gas.

\end{abstract}


\maketitle


Ninety years ago, Fermi derived the thermodynamics of a gas of particles obeying the Pauli exclusion principle~\cite{Fermi1926a}. The Fermi gas quickly became a ubiquitous paradigm in many-body physics; yet even today, Fermi gases in the presence of strong interactions pose severe challenges to our understanding. Ultracold atomic Fermi gases have emerged as a flexible platform for studying such strongly correlated fermionic systems~\cite{DeMarco1999,Inguscio2008,Giorgini2008,Zwerger2012,Zwierlein2015}. In contrast to traditional solid state systems, quantum gases feature tunable spin polarization, dimensionality, and interaction strength. This enables the separation of quantum statistical effects from interaction-driven effects, and invites the exploration of rich phase diagrams, for example bulk Fermi gases in the BEC-BCS crossover~\cite{Schirotzek2009, Nascimbene2009, Koschorreck2012, Kohstall2012, Giorgini2008,Inguscio2008,Zwerger2012,Zwierlein2015} and Fermi-Hubbard models in optical lattices~\cite{Jordens2008, Schneider2008, Esslinger2010, Bloch2012, Hart2015, Greif2016, Cheuk2016a,  Parsons2016,Boll2016,Cheuk2016}.


So far, Fermi gas experiments have been performed in inhomogeneous traps, where the non-uniform density leads to spatially varying energy and length scales. This poses a fundamental problem for studies of critical phenomena for which the correlation length diverges. Furthermore, in a gas with spatially varying density, a large region of the phase diagram is traversed, potentially obscuring exotic phases that are predicted to occur in a narrow range of parameters~\cite{Radzihovsky2010}. A prominent example is the FFLO state~\cite{Larkin1964,Fulde1964}, an elusive supersolid where spin-imbalance imposes a characteristic spatial period that is well-defined only in a homogeneous setting. A natural solution to these problems is the use of uniform potentials, which have recently proved to be advantageous for thermodynamic and coherence measurements with Bose gases~\cite{Gaunt2013, Schmidutz2014, Navon2015, Chomaz2015}.  

Here, we realize homogeneous Fermi gases in a versatile uniform potential. For spin-polarized gases, we observe both the formation of the Fermi surface and the saturation at one fermion per momentum state, due to Pauli blocking. Spatially uniform pair condensates are observed for spin-balanced gases, offering strong prospects for the study of spin imbalanced superfluidity.

\begin{figure}[!t]
\centering
\includegraphics[width=8.6cm]{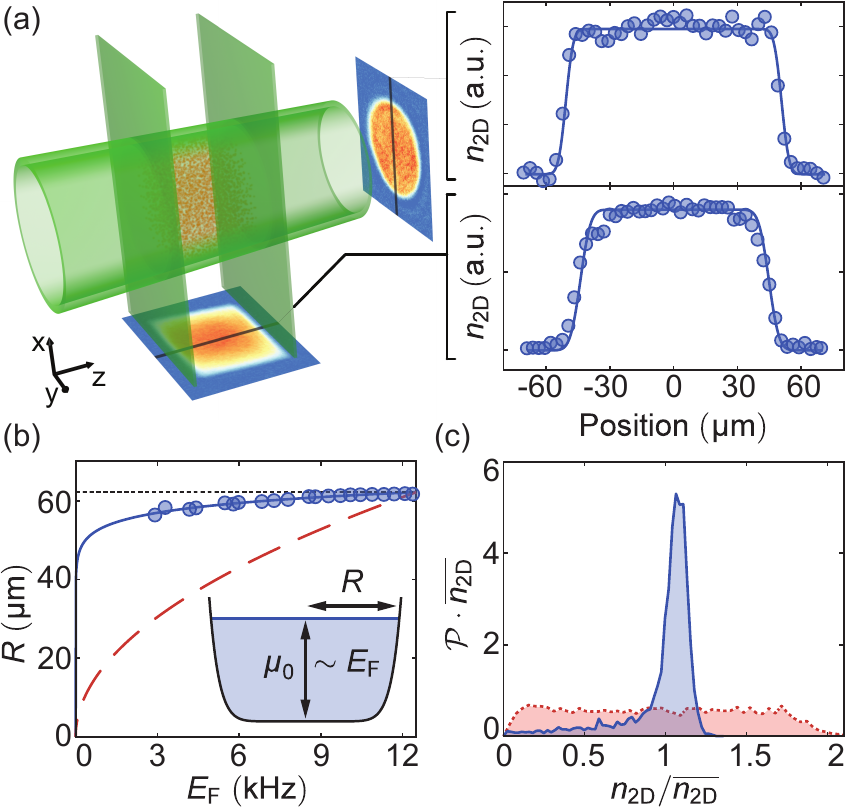}
\caption{(Color online) Homogeneous Fermi gas. (a) Schematic of the box trap and cuts through the column-integrated density profiles along the axial and radial directions. (b) Radius of the cloud as a function of the Fermi energy. The dotted black and dashed red lines correspond to a perfect box potential and a harmonic potential respectively, and are scaled to converge at the highest $E_\mathrm{F}$. The blue solid line corresponds to a power law potential $V(r)\,{\sim}\,r^{16}$. (c) Measured radial probability density $\mathcal{P}(n_{2\mathrm{D}})$  for the column-integrated density $n_{2\mathrm{D}}$, averaging about 20 in-trap images. The blue solid and red dashed lines correspond to the uniform and gaussian traps, respectively.}
\label{fig:M1}
\end{figure}

\begin{figure*}
\centering
\includegraphics[width=\textwidth]{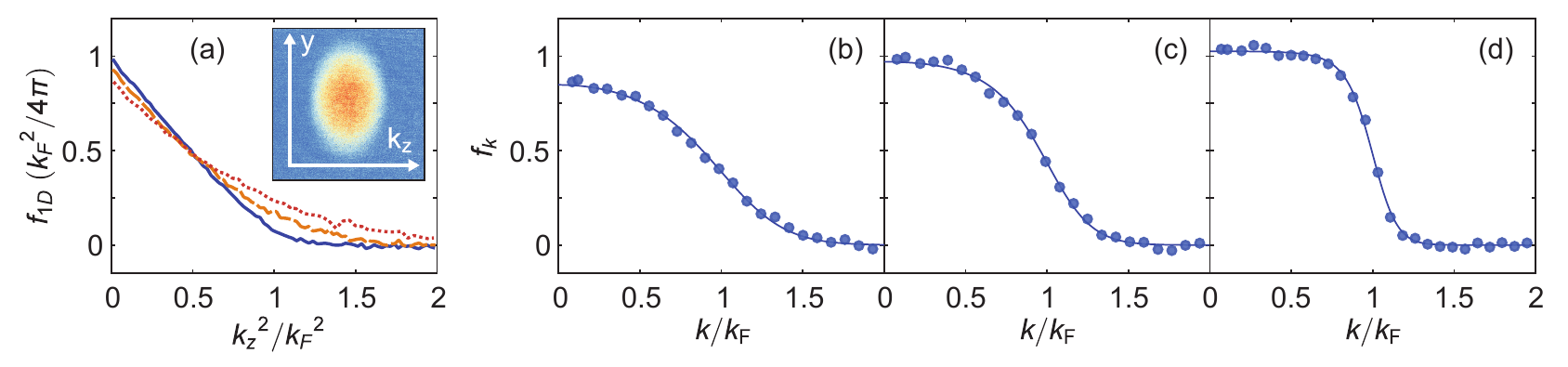}
\caption{(Color online) Momentum distribution of the homogeneous spin-polarized Fermi gas. (a) Doubly integrated momentum distribution ${f}_{1\mathrm{D}}({k_{z}}^2)$ for different temperatures in the uniform trap. In order of decreasing temperature: red dotted line, orange dashed line, and blue solid line. Each line corresponds to averages over 7 images. The optical density after momentum space mapping along $z$ is shown in the inset.  (b), (c), (d): Momentum distribution $f_k = -4\pi \mathrm{d}f_{1\mathrm{D}}/\mathrm{d}{k}^2$, showing Pauli blocking and Fermi surface formation. Fermi-Dirac fits (solid line) give:  (b) $T/T_{F}=0.49(2)$, (c) $T/T_{F}=0.32(1)$, and (d) $T/T_{F}=0.16(1)$, with $k_\mathrm{F}$ ranging between $2.8\,\mathrm{\mu m}^{-1}$ and $ 3.7\,\mathrm{\mu m}^{-1}$. The estimated systematic error in the measurement of $f_k$ is 15\%.}
\label{fig:M2}
\end{figure*}


In cases where the local density approximation (LDA) is valid, the spatially varying local chemical potential in an inhomogeneous trap can be utilized for thermodynamic~\cite{Ho2009,Nascimbene2010,Navon2010,Ku2012} and spectroscopic~\cite{Shin2007, Schirotzek2008, Schirotzek2009} measurements. However, reconstructing the local density from line-of-sight integrated density profiles typically increases noise, while spatially selecting a central region of the gas reduces signal. A potential that is uniform along the line-of-sight is the natural solution. Combining the desirable features of homogeneous and spatially varying potentials, we introduce a hybrid potential that is uniform in two dimensions and harmonic in the third. The line-of-sight integration is now turned into an advantage: instead of averaging over a wide region of the phase diagram, the integration yields a higher signal-to-noise measurement of the local density. Using this geometry, we observe the characteristic saturation of isothermal compressibility in a spin-polarized gas, while a strongly interacting spin-balanced gas features a peak in the compressibility near the superfluid transition~\cite{Ku2012}.


In our experiment, we prepare atoms in the two lowest hyperfine states of~\Li near a Feshbach resonance, and load them into the uniform potential of the optical box trap depicted in Fig.~\ref{fig:M1}(a), after evaporative precooling in a crossed dipole trap. We typically achieve densities and Fermi energies of up to $n\,{\approx}\,10^{12}  \, \mathrm{cm}^{-3}$ and $E_{\mathrm{F}}\,{\approx}\,h \,{\cdot}\,13 \,\mathrm{kHz}$, corresponding to ${\sim}10^6$ atoms per spin state in the box. The lifetime of the Fermi gas in the box trap is several tens of seconds. The uniform potential is tailored using blue-detuned laser light for the confining walls. The sharp radial trap barrier is provided by a ring beam generated by an axicon~\cite{McLeod1954,Manek1998}, while two light sheets act as end caps for the axial trapping~\cite{suppmat}. Furthermore, the atoms are levitated against gravity by a magnetic saddle potential~\cite{Inguscio2008}. The residual radial anti-confining curvature of the magnetic potential is compensated optically, while an axial curvature results in a weak harmonic potential described by a trapping frequency of $\omega_\mathrm{z} = 2\pi \cdot 23.9 \, \mathrm{Hz}$. This typically results in a variation of the potential along the axial direction that is less than $5\%$ of the Fermi energy. We characterize the steepness of the trap walls by measuring the radial extent $R$ of the cloud as a function of Fermi energy (see Fig.~\ref{fig:M1}(b)). Modeling the trap walls with a power law potential, we obtain $V(r) \sim r^{16.2 \pm 1.6}$~\cite{suppmat}.

A stringent measure of the homogeneity of the gas is the probability distribution $\mathcal{P}(n)$ for the atomic density $n$. Imaging along the $z$ and $x$ directions yields the radial and axial probability distribution $\mathcal{P}(n_{2\mathrm{D}})$ for the column density $n_{2\mathrm{D}}$ (see Fig.~\ref{fig:M1}(c) and ~\cite{suppmat}). The distribution for the homogeneous gas is sharply peaked near the trap average density $\overline{n_\mathrm{2D}}$. For comparison, we also show $\mathcal{P}(n_{2\mathrm{D}})$ for an optical gaussian trap, which is spread over a large range of densities.

 
 Fermions at low temperatures are characterized by Pauli blocking~\cite{Fermi1926a}, resulting in the emergence of a Fermi surface near the Fermi wavevector $k_{\mathrm{F}}$ and the saturated occupation of momentum states below $k_{\mathrm{F}}$. Consequences of Pauli blocking have been observed in ultracold gases, for example Pauli pressure~\cite{Truscott2001, Schreck2001}, reduced collisions~\cite{DeMarco2001, Sommer2011}, anti-bunching in noise correlations~\cite{Rom2006}, and the reduction of density fluctuations~\cite{Muller2010,Sanner2010}. In optical lattices under microscopes, Pauli blocking has been observed in real space through observations of band insulating states~\cite{Omran2015, Greif2016, Cheuk2016a} and of the Pauli hole in pair correlations~\cite{Cheuk2016}. Typically obscured in the time of flight expansion of an inhomogeneous atomic gas, the Fermi surface has been observed by probing only the central region of a harmonically trapped gas~\cite{Drake2012}. The uniform box potential enables us to probe the momentum distribution of a homogeneous spin-polarized Fermi gas, and directly observe Pauli blocking in momentum space.

To measure the momentum distribution $f(\vect{k})$, we release a highly spin-imbalanced gas $(n_{\downarrow}/n_{\uparrow}<0.05$, where $n_{\uparrow}$ and $n_{\downarrow}$ are the densities of the majority and minority spin-components, respectively) from the uniform potential into the small residual axial harmonic potential (along the $z$-axis). To ensure the ballistic expansion of the gas, the minority component is optically pumped into a weakly interacting state within $5 \, \mathrm{\mu s}$ ~\cite{suppmat}. After a quarter period of expansion in the harmonic trap, the axial momenta $k_z$ are mapped into real space via $z = \hbar k_z/m\omega_z$ ~\cite{Shvarchuck2002,VanAmerongen2008,Tung2010,Murthy2014}. In contrast to conventional time of flight measurements, this method is unaffected by the in-trap size of the gas. The measured integrated density profile $n_{\mathrm{1D}}(z) = \iint \mathrm{d}x \, \mathrm{d}y \, n(x,y,z)$ reflects the integrated momentum distribution $f_{1\mathrm{D}}(k_{z}) = (2\pi)^{-2}\iint \mathrm{d}k_{x} \, \mathrm{d}k_{y} \, f(k_{x},k_{y},k_{z}) $ via
\begin{equation}
f_{1\mathrm{D}}(k_{z}) = \frac{2\pi \hbar }{V m \omega_{z}} \, n_{\mathrm{1D}}(z).
\end{equation}
Here, $V$ is the volume of the uniform trap. Figure~\ref{fig:M2}(a) shows the integrated momentum distribution for different temperatures. Assuming a spherically symmetric momentum distribution, $f_k\equiv f(\mathbf{k}) = f(k^2)$.  Noting that $\int {\rm d}k_x k_y f(k_x^2+k_y^2+k_z^2) = \pi \int_{k_z^2}^\infty {\rm d} (k^2) f(k^2)$, the three-dimensional momentum distribution can be obtained from the integrated momentum distribution by differentiation:
\begin{equation}
f_k = -4\pi \frac{\mathrm{d}f_{1\mathrm{D}}(k^2)}{\mathrm{d}{k}^2}.
\end{equation}
As the temperature is lowered, the momentum distribution develops a Fermi surface, and we observe a momentum state occupation of $1.04(15)$ at low momenta (see Fig.~\ref{fig:M2}(b)-(d)), where the error in $f_k$ is dominated by the systematic uncertainties in the box volume and the imaging magnification~\cite{suppmat}. This confirms Pauli blocking in momentum space.

\begin{figure}[!t]
\centering
\includegraphics[width=8.6cm]{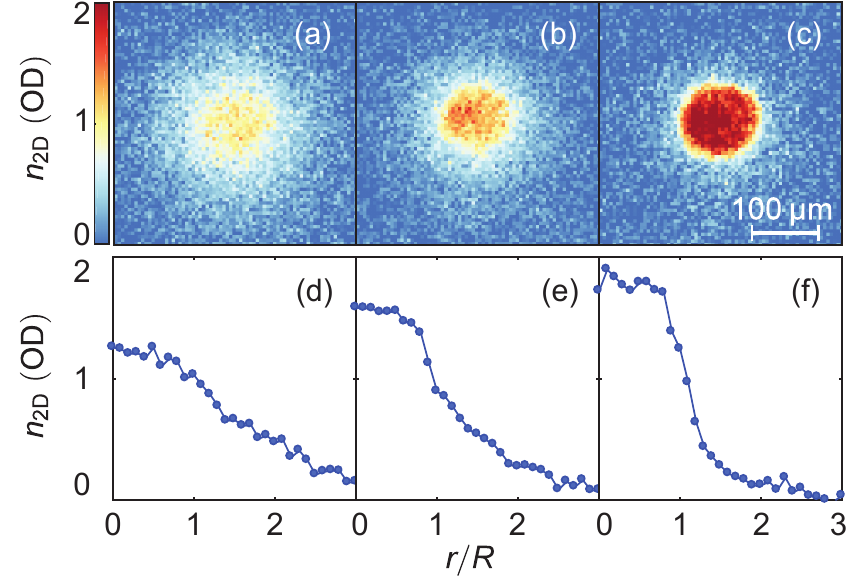}
\caption{(Color online) Pair condensation in the uniform trap. (a), (b) and (c): absorption images after a rapid ramp of the magnetic field and 10 ms of time of flight. The temperature of the gas is lowered (left to right) by evaporation in the uniform trap. The onset of a bimodal distribution signals the formation of a pair condensate. (d), (e) and (f) show cuts through the images in the top row.}
\label{fig:M3}
\end{figure}


An important motivation for the realization of a homogeneous Fermi gas is the prospect of observing exotic strongly correlated states predicted to exist in narrow parts of the phase diagram, such as the FFLO state~\cite{Larkin1964,Fulde1964}. In a harmonic trap, such states would be confined to thin iso-potential shells of the cloud, making them challenging to observe. We observe pair condensation in a uniformly trapped strongly interacting spin-balanced Fermi gas through a rapid ramp of the magnetic field during time of flight \cite{Regal2004,Zwierlein2004,Inguscio2008}, as shown in Fig.~\ref{fig:M3}(a)-(c). The pair condensate at the end of the ramp barely expands in time of flight. As a result, the in-trap homogeneity is reflected in a flat top profile of the condensate (see Fig.~\ref{fig:M3}(f)).

\begin{figure}
\centering
\includegraphics[width=8.6cm]{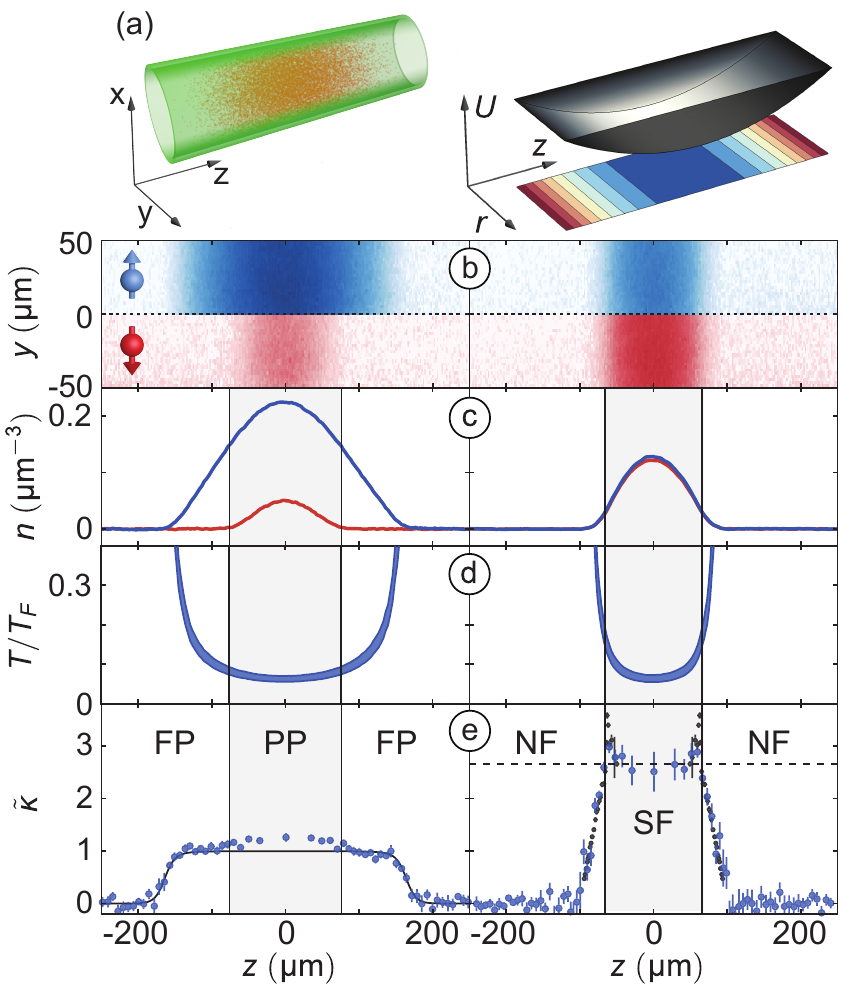}
\caption{(Color online) Unitary Fermi gases in the hybrid trap. (a) Schematic and potential of the trap. The cloud is imaged along an equipotential direction ($x$-axis). Left panels of (b)-(e) show a spin-polarized gas above the Clogston-Chandrasekhar limit, whereas the right side corresponds to a spin-balanced gas. The data is averaged over 6 images. (b) local density obtained from column density for both spin components. (c) Average density for each $x-y$ equipotential slice. The blue (red) line shows the spin-up (-down) component. (d) Spatially resolved temperature of the gas. The blue shaded region represents the error in the temperature determination. (e) Compressibility of the gas. The solid line in the left panel is the compressibility for an ideal Fermi gas. The crossover from fully polarized (FP) region to partially polarized (PP) region is accompanied by an increase in $\widetilde{\kappa}$. Black data points in the right panel correspond to a precision measurement of the balanced unitary equation of state in harmonic trap \cite{Ku2012}. The peaks in the compressibility signal the  phase transition from normal (N) to superfluid (SF). The horizontal dashed line shows the zero-temperature equation of state $\kappa/\kappa_0=1/\xi$.}
\label{fig:M4}
\end{figure}

Although a fully uniform potential is ideal for measurements that require translational symmetry, a spatially-varying potential can access a large region of the phase diagram in a single experimental run. To harness the advantages of both potentials, we introduce a hybrid geometry that combines the radially uniform cylinder trap with an axially harmonic magnetic trap along the $z$-direction (see Fig.~\ref{fig:M4}(a)). As a benchmark for the hybrid trap, we perform a thermodynamic study of both a spin-polarized and a spin-balanced unitary gas. Figure \ref{fig:M4}(c)-(e) display for both cases the $y$-axis averaged local density, temperature, and compressibility. The data shown in Fig.~\ref{fig:M4} is extracted from an average of just six images per spin-component. For comparison, precision measurements of the equation of state at unitarity, performed in conventional harmonic traps, required averaging of over $100$ absorption images~\cite{Ku2012}. The temperature is obtained from fits to the known equations of state of the non-interacting and spin-balanced unitary Fermi gas respectively. 
From the local density in the hybrid trap, we determine the normalized isothermal compressibility $\widetilde{\kappa} = \frac{\kappa}{\kappa_0}= - \left. \frac{\partial E_{\mathrm{F}}}{{\partial U}}  \right| _{T}$ for the spin-polarized and the spin-balanced gas. Here, $U$ is the external potential, and $\kappa_0=\frac{3}{2}\frac{1}{n E_{\mathrm{F}}}$ is the compressibility of the non-interacting Fermi gas at zero temperature~\cite{Ku2012}.

The spin-polarized cloud features two distinct regions in the trap. The center of the cloud is a partially polarized region in which $(n_{\uparrow}-n_{\downarrow})/(n_{\uparrow}+n_{\downarrow})>0.64$, well above the Clogston-Chandrasekhar limit of superfluidity~\cite{Chandrasekhar1962, Clogston1962, Zwierlein2006}. Surrounding the center is a fully polarized region, where the compressibility is seen to saturate: the real space consequence of the Pauli blocking in momentum space demonstrated in Fig.~\ref{fig:M2}. In contrast, for a non-interacting Bose gas, the compressibility would diverge for $T \rightarrow 0$.  

The majority spin component in the partially polarized region is affected by the presence of the minority spin component. We measure the compressibility $\widetilde{\kappa}_{\uparrow} = - \frac{\partial E_{\mathrm{F}\uparrow}}{{\partial U}}$ in the partially polarized region, and observe an increase compared to the fully polarized gas. This is expected as the minority atoms in the center of the trap attract majority atoms and form polarons~\cite{Schirotzek2009, Nascimbene2009}. The effect is indeed predicted by the polaron equation of state~\cite{Shin2008b, Nascimbene2010, Navon2010}. The observation of this subtle effect highlights the sensitivity of the hybrid potential for thermodynamic measurements.

In the spin-balanced case, $\kappa/\kappa_0$ is significantly larger than for the ideal Fermi gas due to strong interactions. The two prominent peaks in the reduced compressibility signal the superfluid transition at the two boundary surfaces between the superfluid core and the surrounding normal fluid. Near the center of the trap, the reduced compressibility agrees with the $T=0$ equation of state $\kappa/\kappa_0=1/\xi=2.65(4)$, where $\xi$ is the Bertsch parameter. The shaded region in the right column of Fig.~\ref{fig:M4} shows the superfluid part of the gas, where the temperature is below the critical temperature for superfluidity $T_\mathrm{c}=0.17\,T_{\mathrm{F}}$~\cite{Ku2012}.

The realization of uniform Fermi gases promises further insight into phases and states of matter that have eluded observation or quantitive understanding. This includes the observation of the quasiparticle jump~\cite{Mahan2000} in the momentum distribution of a Fermi liquid, critical fluctuations in the BEC-BCS crossover, and long lived solitons \cite{Ku2016}. Of particular interest are spin imbalanced mixtures that have been studied extensively in harmonic traps \cite{Partridge2006,Zwierlein2006,Shin2006,Shin2008,Nascimbene2010,Navon2010,Liao2010}, where the trap drives the separation of normal and superfluid phases into a shell structure. This phase separation should occur spontaneously in a uniform spin-imbalanced gas, possibly forming domains of superfluid and eventually ordering into an FFLO state. In addition, the hybrid potential is a valuable tool for precision measurements that rely on an in-trap density variation. For example, spatially resolved RF spectroscopy~\cite{Shin2007} in the hybrid potential would measure the homogenous response of the system over a large range of normalized temperatures $T/T_\mathrm{F}$ in a single experimental run.

This work was supported by the NSF, the ARO MURI on Atomtronics, AFOSR PECASE and MURI on Exotic Phases, and the David and Lucile Packard Foundation. We would like to thank R. Fletcher for a critical reading of the manuscript, and E. A. Cornell and C. \mbox{Altimiras} for helpful discussions. Z.H. acknowledges support from EPSRC [Grant No. EP/N011759/1].

\clearpage


\section*{Supplementary material}

# Supplemental Material:
# Homogeneous Atomic Fermi Gases

Biswaroop Mukherjee,[1] Zhenjie Yan,[1] Parth B. Patel,[1]
Zoran Hadzibabic,[1,2] Tarik Yefsah,[1,3] Julian Struck,[1] and Martin W. Zwierlein[1]

[1] MIT-Harvard Center for Ultracold Atoms, Research Laboratory of Electronics, and Department
of Physics, Massachusetts Institute of Technology, Cambridge, Massachusetts 02139, USA

[2] Cavendish Laboratory, University of Cambridge, J. J. Thomson Avenue, Cambridge CB3 0HE, United Kingdom

[3] Laboratoire Kastler Brossel, CNRS, ENS-PSL Research University,
UPMC-Sorbonne Universités and Collège de France, Paris, France

## CYLINDER-SHAPED TRAP

For the repulsive optical potential, we use laser light that is blue detuned with respect to the D-lines of $^6$Li at 671 nm. The laser source is a multi-mode 10W laser at 532 nm. Figure S1 shows the optical setup that is used to shape the beam into a hollow core cylinder. A collimated gaussian beam propagates through an axicon and a microscope objective, generating a hollow cylindrical beam in the Fourier plane [S2, S3]. An opaque circular silver mask is placed in the focal plane to block the residual light inside the ring and provide a sharper inner edge. The resulting intensity distribution at the focal plane is imaged onto the atoms along the z-axis. This confines the atoms into a cylinder oriented along the axial direction (z-axis).

In addition to the radial cylinder-shaped trap, the uniform trap requires sharp end cap walls that confine the atoms along the axial direction. For the endcaps, we use second 532 nm beam from the same laser source and detuned it by 160 MHz to avoid interference between the beams. The end cap beam is split into two elliptically shaped beams with opposite polarizations, which are fo-

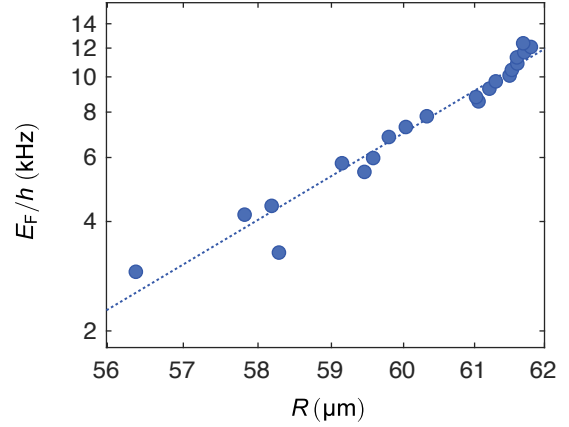

FIG. S2. Determination of the power law exponent of the radial wall potential. Log-log plot of the data shown in Fig. 1(b). The blue dotted line is a linear fit, with a slope of $m = 16.2 \pm 1.6$ for the power law exponent.

cused onto the edges of a rectangular opaque mask. The intensity distribution at the mask is projected onto the atoms and provides two sharp confining walls.

## TRAP CHARACTERIZATION

### Radial Trap Wall: Power Law Potential

To describe the radial extent of the gas as a function of the Fermi energy, we model our radial potential with a power law $U(r) = \alpha r^m$. Within the local density approximation, the local chemical potential is then determined by $\mu(r) = \mu_0 - \alpha r^m$, where $\mu_0 = \mu(r = 0)$. The radius measurements have been performed with a spin-balanced superfluid. Assuming $T = 0$, the cloud has a well defined Thomas-Fermi radius $R$, where the density drops to zero:

$$\mu_0 = \xi E_F = \alpha R^m, \tag{S1}$$

with the Bertsch parameter $\xi$. Fitting the data shown in Fig. S2 with a power-law gives $m = 16.2 \pm 1.6$.

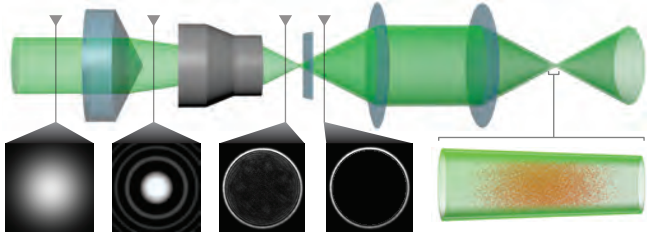

FIG. S1. Optical setup for cylinder-shaped trap. From left to right: A gaussian beam propagates through an axicon resulting in a Bessel beam in the near field. Subsequently the Bessel beam is focused through a microscope objective. In the focal plane, the resulting intensity pattern is a ring with gaussian rim. A matched circular opaque mask is used to block out residual light in the center of the ring. Finally the mask is projected through an imaging system onto the atoms, creating the cylinder-shaped trap for the atoms. A small variation of cylinder radius is unavoidable when using a single axicon [S1].

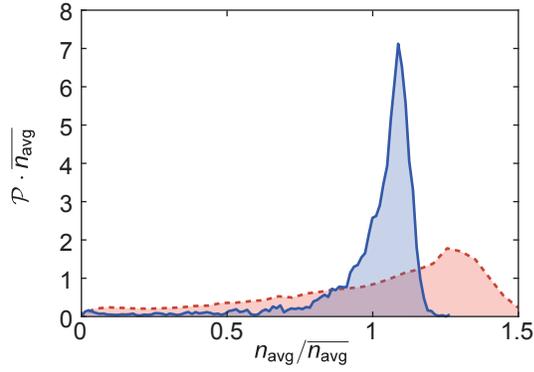

FIG. S3. Axial probability distribution. The blue solid line (red dashed line) shows $\mathcal{P}(n_{avg})$ for the uniform (hybrid) trap. Absorption images are taken along x-axis. The line of sight averaged local density $n_{avg}$ is calculated assuming a uniform cylindrical trap.

## Axial Probability Distribution

To extract information about the axial homogeneity of the gas, we image the atoms along the x-axis. We obtain the line-of-sight averaged local density $n_{avg}$ from the column density by dividing by the local column length. Figure S3 shows the probability distribution $\mathcal{P}(n_{avg})$ for the line-of-sight averaged density of the uniform and hybrid trap. The axial probability distribution shows a narrow peak similar to the one observed for the radial distribution. The probability distribution for the hybrid trap is broadened due to the harmonic trapping along the z direction.

## OPTICAL PUMPING OF THE MINORITY ATOMS FOR MOMENTUM-SPACE MAPPING

The measurement of the momentum distribution relies on ballistic expansion of the gas immediately after release from the trap. However, the expansion of the atoms is strongly influenced by a small minority fraction ($< 5\%$) of strongly interacting atoms admixed to ensure the thermalization of the gas. To eliminate the interactions between the two spin states during the expansion, the minority atoms are optically pumped into the hyperfine state $|m_J = +1/2, m_I = 0\rangle$ that is weakly interacting with the majority cloud. The $5\,\mu s$ pumping pulse is applied right before the release of the atoms into the harmonic trap. On average, 1.5 photons are required to pump an atom into the weakly interacting hyperfine state. The transitions involved in this pumping scheme are shown in the inset of Fig. S4. Figure S4 shows the integrated momentum distribution of the gas obtained using momentum-space mapping with and without pumping of minority atoms. Note that without the optical pumping of the minority atoms, $f_{1D}(k_z^2)$ is distorted

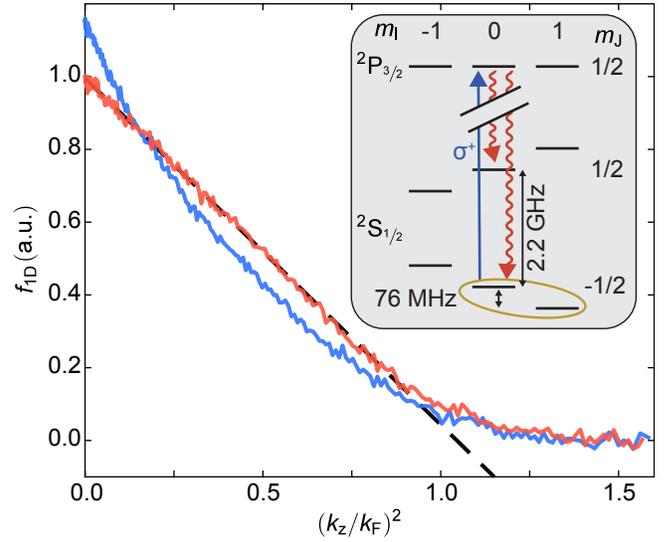

FIG. S4. Influence of the minority atoms on the expansion of the majority cloud. The blue (red) line corresponds to the observed integrated momentum distribution without (with) optical pumping of the minority atoms. The dashed line is a guide to the eye. The inset displays the level scheme for the optical pumping at a magnetic field of $B = 832\,G$. The yellow ellipse marks the two strongly interacting spin states, where the minority is in $|m_J = -1/2, m_I = 0\rangle$. The pumping transition is shown as the blue line and the spontaneous decay channels with red curvy lines.

from the triangular shape expected for a low temperature Fermi gas.

## DENSITY MEASUREMENT WITH ABSORPTION IMAGING

For heavy atoms, such as Rb and Cs, and imaging intensities that are small compared to the saturation intensity ($I_{sat}$), the column density $n_{col} = -(1/\sigma_0)\log(I_f/I_i)$ is determined by the Beer-Lambert law. Here, $I_i$ and $I_f$ are the intensities of the imaging beam before and after the atoms, respectively, and $\sigma_0$ is the absorption cross-section. However, for light atoms such as Li, the Doppler effect plays a dominant role in realistic experimental scenarios, where a low imaging intensity and short exposure time is in conflict with a high signal to noise ratio. For our experiment, depending on the column density of the sample, preferred values for the imaging intensities are $0.1 - 0.5\,I_{sat}$ at an exposure time of $4 - 10\,\mu s$. Under these conditions, each $^6Li$ atom scatters up to 35 photons. The corresponding photon recoil results in a Doppler shift of up to 6 MHz, which is comparable to the natural linewidth of $^6Li$.

In order to account for the Doppler and saturation effects, we numerically solve two coupled differential equations for the local, time-dependent saturation parameter

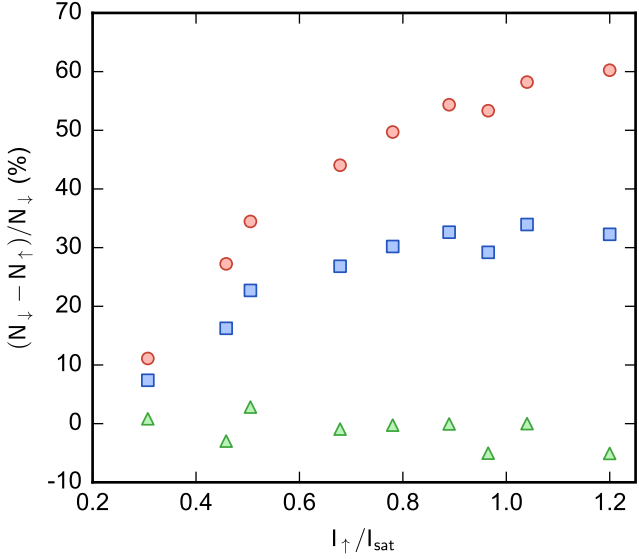

FIG. S5. Apparent atom count for the spin up component ($N_\uparrow$) measured at different imaging intensities ($I_\uparrow$). A reference atom count ($N_\downarrow$) is obtained by subsequently imaging a second spin component at a fixed imaging intensity ($I_\downarrow = 0.23 I_{sat}$). Red circles, blue squares, and green triangles are obtained using Beer-Lambert, saturated Beer-Lambert, and Doppler Beer-Lambert, respectively. Spin balanced clouds are used for these measurements.

$s(z,t) = I(z,t)/I_{sat}$ and velocity $v(z,t)$;

$$\frac{\partial s}{\partial z} = -n\sigma_0 \frac{s}{1 + s + (2kv/\Gamma)^2}. \tag{S2a}$$

$$\frac{\partial v}{\partial t} = \frac{\hbar k \Gamma}{2m} \frac{s}{1 + s + (2kv/\Gamma)^2}. \tag{S2b}$$

Here, $\sigma_0$, $k$, $m$ and $\Gamma$ are the bare scattering cross-section, photon wave vector, atomic mass and natural linewidth.

In the limit where the Doppler effect is negligible ($v = 0$), the analytical solution of Eq. S2 is $n_{col}\sigma_0 = -\log(s_f/s_i) - (s_f - s_i)$. Here, $s_i$ and $s_f$ are reduced imaging intensities before and after the atoms, respectively. This is the modified version of the Beer-Lambert law that includes saturation of the atomic transition. For the general case, we numerically solve Eq. S2 to find the Doppler-corrected relation between $I_i$ and $I_f$ (which we call the "Doppler Beer-Lambert" law).

We compare these aforementioned methods by subsequently imaging the two spin states of a spin-balanced gas with a fast SCMOS camera. The first image is taken with fixed saturation intensity of $s_\downarrow = 0.23$ and serves as the density reference. The second absorption image for the other spin component, with a variable $s_\uparrow$, is obtained $15\,\mu s$ after the first image. Figure S5 shows the differences in the measured total atom numbers between two spin components for various $s_\uparrow$ calculated using the Beer-Lambert law (red circles), the saturated Beer-Lambert law (blue squares) and Doppler Beer-Lambert (green triangles). For the atom number differences calculated using the Doppler Beer-Lambert law, the mean deviation from the reference density is only 3% compared to 27% and 46% for the saturated Beer-Lambert and the basic Beer-Lambert law, respectively.

---



\end{document}